\documentclass[12pt,a4paper]{article}
\usepackage{psfig}

\newcommand{\be}{\begin{equation}}
\newcommand{\ee}{\end{equation}}
\newcommand{\bel}[1]{\be\label{#1}}

\newcommand{\ber}{\begin{eqnarray}}
\newcommand{\eer}{\end{eqnarray}}

\newcommand{\psib}{\overline{\psi}}
\newcommand{\goo}{\,\raisebox{-.5ex}{$\stackrel{>}{\scriptstyle\sim}$}\,}

\newcommand{\dd}{\partial\hspace{-7pt}/}
\newcommand{\hsp}{\hspace*{1pt}}

\begin{document}
\vspace*{-1cm}
\begin{center}
{\large\bf Catastrophic rearrangement of a compact star\\
due to the quark core formation}\\[5mm]

{I.N. Mishustin$^{\,1,2,3}$, M. Hanauske$^{\,1}$, 
A. Bhattacharyya$^{\,1,4}$, L.M. Satarov$^{\,1,2}$,\\ 
H. St\"ocker$^{\,1}$, and W. Greiner$^{\,1}$}
\end{center}

\begin{tabbing}
\hspace*{1cm}\=
${}^1$\={\it Institut f\"ur Theoretische Physik,
J.W.~Goethe--Universit\"at,}\\ 
\>\>{\it D--60054~Frankfurt~am~Main,~Germany}\\
\>${}^2$\>{\it The Kurchatov Institute, Russian Research
Center, 123182 Moscow, Russia}\\
\>$^3$\>{\it The Niels Bohr Institute, DK-2100 Copenhagen {\O}, Denmark}\\
\>${}^4$\>{\it Scottish Church College, 1\&3  Urquhart Square,
700 006 Calcutta, India}
\end{tabbing}

\begin{abstract}  
We study properties of compact stars with  the deconfinement
phase transition in their interiors.  The equation of state of cold baryon-rich
matter is constructed by combining a relativistic mean-field model for the
hadronic phase and the MIT Bag model for the deconfined phase. In a narrow
parameter range two sequences of compact stars (twin stars), which differ by
the size of the quark core, have been found.  We demonstrate the possibility of
a rapid transition between the twin stars with the energy release of about
$10^{52}$ ergs. This transition should be accompanied by the prompt neutrino
burst and the delayed gamma-ray burst.  
\end{abstract}
\hspace*{15mm}PACS: 14.65.-q, 26.60.+c, 97.10.-q, 98.70.Rz

\section{Introduction}

Our main goal in this paper is to study properties of compact stars 
composed of strongly interacting matter undergoing the deconfinement 
phase transition. We construct the equation of state (EoS) at finite 
baryon density $\rho_B$ by combining two popular models of the
baryon-rich matter. For the hadronic phase we use the NLZ version of 
the relativistic mean-field model \cite{rufa88} which gives very good 
description of the saturation properties of cold nuclear matter. 
The deconfined phase is described within a simplified version of the 
MIT Bag model~\cite{cho74}. The possibility of a phase  transition between
these two phases has been carefully studied  by applying the Gibbs conditions
for charge-neutral and \mbox{$\beta$-equilibrated} matter. Indeed, we have
found a first order phase transition and determined characteristics of the
mixed phase.

Properties of hybrid stars containing both the hadrons and quarks have 
been already studied by using a large number of models (see e.g.
Refs.~\cite{gle97,stei00,han01,scha02,bur02}). One may wonder, what is new in
our work? First of all, we want to note that this paper presents only a small
part of our more comprehensive study~\cite{han02} where we systematically 
analyze many other
models of hadronic and quark phases. It is interesting that in most cases we do
not find any phase transition between the two phases. The combination of the
NLZ and MIT Bag models represents one of a few exceptional cases, when the
deconfinement  phase transition is predicted in stellar interiors. Moreover, 
we find two families of compact stars, twin stars~\cite{ger68,gle98,sche00},
which differ by the size of the quark core. This opens the possibility of a
catastrophic rearrangement of the twin star from one to the other configuration
with a release of energy of  about 10$^{52}$ erg. In this paper we present
these results and  discuss their observational consequences.

\section{Properties of matter in compact stars} 

\subsection{Hadronic phase}

There is no doubt that hadrons, mesons and baryons, are correct degrees of
freedom for modeling strongly-interacting matter at low densities. From nuclear
phenomenology we know that atomic nuclei can be well described in terms of
interacting nucleons. Therefore, we believe that the hadronic phase should be
stable at least up to the saturation density of nuclear matter~$\rho_0 
= 0.15 \, {\rm {fm^{-3}}}$\,. 
Field-theoretical models, where nucleons interact with mean meson fields, are
proved to be very successful in describing saturation properties of nuclear
matter as well as properties of finite  nuclei. Here, to
calculate the EoS of hadronic matter we use a non-linear version of the
relativistic mean-field model known as the NLZ  model~\cite{rufa88}. Compared
wih the original version it is generalized by including hyperons (Y) and
hyperon-hyperon (YY) interactions, as proposed in  Ref.~\cite{scha96}. We call
this model NLZY.

The original Lagrangian density for the NLZ model (without
the YY interaction) can be written as~\cite{rufa88}
(here and below we use the units $\hbar=c=1$)
\ber
{\cal L}&=&\sum_B \psib_B\left(i\dd-m_B \right) \psi_B + {1 \over 2} 
\partial^\mu \sigma \partial_\mu \sigma  
-\frac{1}{2} m_\sigma^2 \sigma^2
-\frac{a}{3}\sigma^3-\frac{b}{4}\sigma^4 
-{1 \over 4} \omega^{\mu\nu} \omega_{\mu\nu}+  
\frac{1}{2} m_\omega^2 \hsp\omega^\mu\omega_\mu \nonumber\\
&-&{1 \over 4} {\vec\rho}^{\,\mu\nu} {\vec\rho}_{\mu\nu}+ 
{1 \over 2} m_\rho^2 {\vec\rho}^{\mu} {\vec\rho}_{\hsp\mu}
+\sum_B \psib_B \left(g_{\sigma B} \sigma + 
g_{\omega B}\omega^\mu\gamma_\mu 
+ g_\rho{\vec\rho}^{\,\mu}\gamma_\mu {\vec\tau}_B\right)\psi_B\,,
\label{lagr}
\eer
where the sum runs over all the baryons $B$=$p, n, \Lambda,\Sigma^{0,\pm}, 
\Xi^{0,-}$. In the above Lagrangian $\sigma, \omega$ and
${\vec\rho}$ are the iso-scalar scalar $\sigma$, the iso-scalar vector $\omega$
and the isovector vector $\rho$ meson fields respectively. 
In Eq.~(\ref{lagr}) $\omega^{\mu\nu}$ and ${\vec\rho}^{\,\mu\nu}$ denote,
respectively, the field strength tensors for the $\omega$ and 
$\rho$ meson fields.

Originally this model was designed for the nucleonic sector and it failed to
reproduce the observed strong $\Lambda \Lambda$ attractive 
interaction. This defect can be removed by adding two new meson
fields with hidden strangeness,  namely, the iso-scalar scalar $\sigma^*$ and 
the iso-vector vector $\phi$, which couple to hyperons only~\cite{scha96}.
These fields can be identified with the $f_0\hsp (975)$ and $\phi\hsp (1020)$ 
mesons. The corresponding Lagrangian is given by
\ber
{\cal L}^{YY}={1 \over 2} \left(\partial^\mu \sigma^* \partial_\mu \sigma^* 
- m^2_{\sigma^*} \sigma^{*2}\right)  - {1 \over 4} \phi^{\mu\nu}
\phi_{\mu\nu}+{1 \over 2} m_\phi^2 \phi^\mu \phi_\mu 
+ \sum_Y \psib_Y \left(g_{\sigma^* Y} \sigma^* 
+ g_{\phi Y} \phi^\mu\gamma_\mu \right)\psi_Y \nonumber
\eer
where index $Y$ runs over hyperons only. The mean meson fields are 
found from the Euler-Lagrange equations. 

The nucleon coupling constants are chosen from the fit of the finite nuclei
properties. The vector coupling constants of the hyperons are chosen according
to the SU(6) symmetry and the hyperonic scalar coupling constants are chosen to
reproduce the measured values of the corresponding optical potentials.
Below we use the set of model parameters suggested in 
Refs.~\cite{rufa88,scha96}.

Within the mean-field approximation,
the pressure and energy density of static and homogeneous 
baryonic matter can be easily calculated from the above Lagrangian:  
\ber
\epsilon^{H} &=& \frac{1}{2}m_\sigma^2 \sigma^2
+ \frac{b}{3}\sigma^3 + \frac{c}{4}\sigma^4
+ \frac{1}{2}m_{\sigma^*}^2 {\sigma^*}^2
+ \frac{1}{2}m_\omega^2 \omega_0^2 + 
\cr && {}
+ \frac{1}{2}m_\rho^2 \rho_{0,0}^2
+ \frac{1}{2}m_\phi^2 \phi_0^2
+ \sum_{B} \frac{\nu_B}{2\pi^2} \int_0^{k_F^B} dk\hsp k^2
\sqrt{k^2 + {m^*_B}^2}~, \\ \cr
P^{H} &=& - \frac{1}{2}m_\sigma^2 \sigma^2
- \frac{b}{3}\sigma^3 - \frac{c}{4}\sigma^4
- \frac{1}{2}m_{\sigma^*}^2 {\sigma^*}^2
+ \frac{1}{2}m_\omega^2 \omega_0^2 + 
\cr && {}
+ \frac{1}{2}m_\rho^2 \rho_{0,0}^2
+ \frac{1}{2}m_\phi^2 \phi_0^2
+ \sum_{B} \frac{\nu_B}{6\pi^2} \int_0^{k_F^B} dk
\frac{k^4}{\sqrt{k^2 + {m^*_B}^2}}\,,
\eer
where $m^*_B=m_B-g_{\sigma B}\sigma-g_{\sigma^* B}\sigma^*$ is the effective 
mass, $\nu_B$ is the degeneracy factor and $k_F^B=\sqrt{\mu_B^2-{m^*}^2_B}$ 
is the Fermi momentum  of the baryon species $B$.

In order to have a complete description of the $\beta$-equilibrated matter one
should  also include the leptons. In both the hadronic and quark phases their 
contributions to energy density and pressure are given by the well known 
formulae of ideal Fermi gas.

\subsection{Deconfined phase}

As follows from a simple geometrical consideration,  nucleons begin to overlap
at densities  $\rho_B\sim (4\pi r_N^3/3)^{-1}\simeq 3\rho_0$ for  the nucleon
radius $r_N\simeq 0.8$ fm.  Such densities are surely  reached in the interiors
of compact stars. Of course, this argument does  not tell anything about the
character of transition from hadronic to quark-gluon degrees of freedom. Below
we follow the common practice of using two different models for these two
phases. Namely, the deconfined phase is described within a simple version of
the MIT Bag model~\cite{cho74}, considering it as a mixture of free  Fermi
gases of $u,d,s$ quarks in a bag with an additional energy density $B$ (the bag
constant).  Within this model the energy density and pressure of cold 
deconfined matter are written as
\begin{eqnarray}
\epsilon^Q &=& \sum_{f=u,d,s} 
\frac{\nu_f}{2 \pi^2} \int_0^{k_F^f} dk k^2\sqrt{m_f^2 + k^2}+ 
B\,,\label{edec}\\ 
P^Q &=& \sum_{f=u,d,s} \frac{\nu_f}{6\pi^2} 
\int_0^{k_F^f} dk \frac{k^4}{\sqrt{m_f^2 + k^2}}- B\,, 
\label{pdec}
\end{eqnarray}
where $k_F^f=\sqrt{\mu_f^2-m_f^2}$ is the Fermi momentum of quarks  with 
flavor~$f$.  For each flavor we choose the degeneracy factor  
$\nu_f = 2\hsp ({\rm spin}) \times 3\hsp ({\rm color}) = 6$ and take the 
following values of quark masses: $m_u=5$ MeV, $m_d=10$ MeV and $m_s=150$~MeV.

\subsection{Conditions of local equilibrium}

In $\beta$-equilibrium, the chemical potential of any particle species $i$ 
can be expressed as
\bel{chem}
\mu_i = b_i\hsp\mu_b + q_i\hsp\mu_e~,
\ee
where $b_i$ is the baryon number of the species $i$\,,
$q_i$ denotes its charge in units of the electron charge, 
$\mu_b$ and $\mu_e$ are the baryonic and electric chemical potentials,
respectively. Here and below we assume that neutrinos can freely escape
from the star. Eq.~(\ref{chem}) means that only reactions conserving 
charge and baryon number are allowed. On the other hand, 
strangeness is not conserved because strangeness-changing 
reactions are generally much faster than
a characteristic time of the star evolution. Two independent chemical
potentials, $\mu_b$ and $\mu_e$, are found by fixing the baryon 
and electric charge densities:
\begin{equation}
\rho_B=\sum_i b_i \rho_i\,,~~~~\rho_e=\sum_i q_i \rho_i\,,
\end{equation}
where $\rho_i$ is the number density of the particle species $i$\,.
It is obvious that stars must be electrically neutral
on a macroscopic scale, i.e.~\mbox{$\rho_e=0$}. 

We assume that deconfinement is a first order phase 
transition which, in general, should produce a mixed phase (MP)
between the pure hadronic phase (HP) and pure quark phase (QP). 
At zero temperature the MP should follow the Gibbs conditions
\begin{eqnarray}
P^H(\mu_b,\mu_e) &=& P^Q(\mu_b,\mu_e)\,,\\ \label{eq:10}
\mu_b \,=\, \mu_b^{H} &=& \mu_b^{Q}\,,\\
\mu_e \,=\, \mu_e^{H} &=& \mu_e^{Q}\,.  
\end{eqnarray}
According to Eq.~(\ref{chem}), the baryon chemical potential
$\mu_b$ equals the neutron chemical potential $\mu_n$\, and
$\mu_e$ is equal to the electron chemical potential.
At given $\mu_b$ and $\mu_e$, the quark chemical potentials
are found by using the formulae
$\mu_u=(\mu_b-2\hsp\mu_e)/3$ and $\mu_d=\mu_s =(\mu_b+\mu_e)/3$\,.

The volume averaged energy density in the MP can be written as
\begin{eqnarray}
\epsilon &=&(1-\lambda)\hsp\epsilon^H(\mu_b,\mu_e) + 
\lambda\hsp\epsilon^Q(\mu_b,\mu_e)\,,
\label{eq:12}
\end{eqnarray}
where $\lambda=V_Q/V$ is the volume fraction of quark phase. 
In the case of two chemical potentials one can only
construct the MP by adopting a generalized (global) charge neutrality
condition~\cite{gle92}, when the net positive charge of one phase 
is compensated by the negative charge of the other phase:
\be
(1-\lambda)\hsp\rho^H_e(\mu_b,\mu_e)+
\lambda\hsp\rho^Q_e(\mu_b,\mu_e) = 0\,.
\ee 
This condition is assumed in our calculations of hybrid stars presented below.

\begin{figure*}[!ht]
\begin{center}
\centerline{\psfig{file=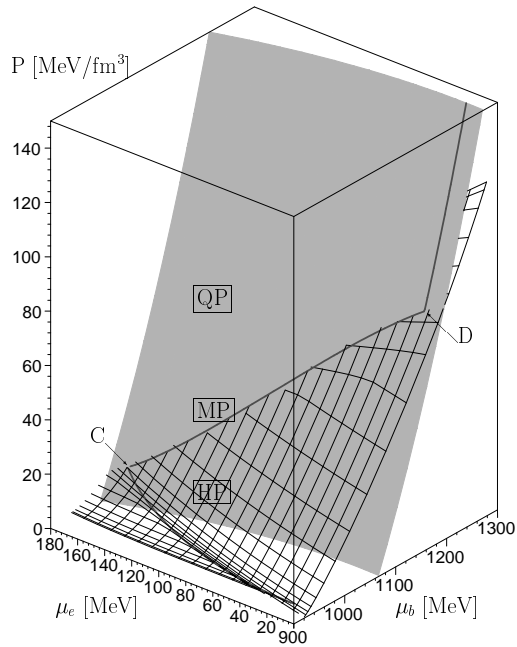,width=0.55\textwidth}}
\vskip-2.0cm
\caption{\label{fig:1}
Pressure as a function of $\mu_b$ and $\mu_e$ 
for the hadronic phase (wired surface) as predicted by the NLZY model 
and for the quark phase (grey surface) as predicted by the MIT Bag model
($B^{1/4} = 180$ MeV). The thick line represents the charge neutral matter. 
Its portion between points C and D corresponds to the mixed phase.}
\end{center}
\vskip-1.25cm
\end{figure*}

\subsection{EoS and composition of matter}

Figure~\ref{fig:1} shows the pressure surfaces for the pure HP and pure QP
(for bag constant $B^{1/4}=180$~MeV)  as functions of chemical potentials
$\mu_b$ and $\mu_e$\,. It is important that these two surfaces intersect, which
is not the case for most of the other models~\cite{han02}. In this case one can
construct a MP connecting the pure phases along the intersection line. The
thick curve on the surfaces shows pressure of charge neutral  matter in 
$\beta$-equilibrium. The lower part of this curve (from low pressure to point
C) corresponds to the pure HP.  The MP starts at point C corresponding to 
baryonic density $\rho_B\simeq 1.5\,\rho_0$\,, and ends at point D
corresponding to $\rho_B\simeq 5.1\,\rho_0$\,.  At  higher densities the matter
is composed purely of quarks. Our calculations show that at $B^{1/4}>190$~MeV
the pressure surfaces of the HP and QP do not intersect at all and the MP can
not be defined by the Gibbs rules. 

\begin{figure*}[!ht]
\begin{center}
\vskip-1.0cm
\centerline{\psfig{file=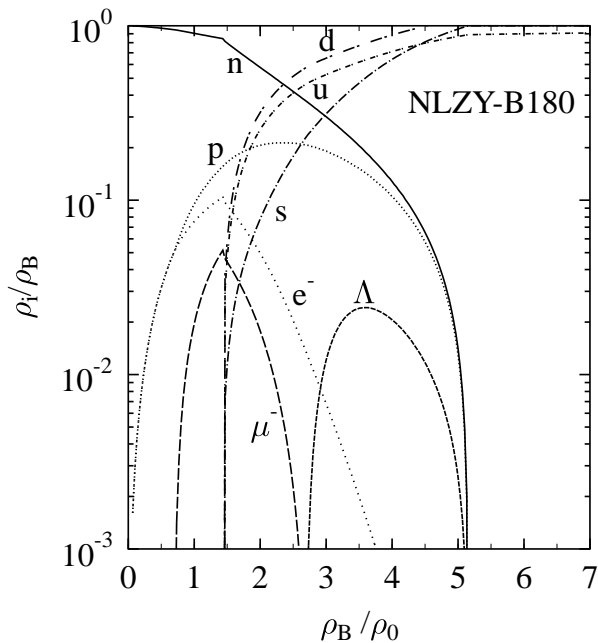,width=0.55\textwidth}}
\vskip-0.8cm
\caption{\label{fig:2}
Relative particle abundances versus baryon density
$\rho_B$ as predicted by the combined NLZY and MIT Bag
($B^{1/4}=180$ MeV) models. The region  \mbox{$1.5<\rho_B/\rho_0<5.1$} 
corresponds to the mixed phase.}
\end{center}
\vskip-1.25cm
\end{figure*}

The particle composition in these three regions is  presented in
Fig.~\ref{fig:2}. It is interesting to note that the only hyperon
species which survive in the MP is $\Lambda$-particle which appears at
densities \mbox{$2.8\hsp\rho_0<\rho_B<5.1\hsp\rho_0$}. All the negatively
charged hyperons are suppressed as a result of the charge neutrality. The
strangeness content of matter rapidly grows in the MP and reaches the asymptotic
value of $1/3$ in the~QP. On the other hand, leptons (mainly electrons) are
practically extinguished at $\rho_B\goo 3\rho_0$\,.

\section{Properties of compact stars}

Below we study properties of static spherically symmetric stars. Under
assumption that the matter may be treated as an ideal fluid, the star structure
can be found by solving the TOV equations~\cite{tol39}. For a given EoS,
$P=P(\epsilon)$\,, and a fixed central baryon  density $\rho_c=\rho_B (r=0)$ we
integrate the TOV equations  from the center of the star up to its surface
$r=R$. The star radius $R$ is determined from the condition $P(R)=0$. The
details of numerical calculation as well as the online program  can be found on
the internet home page~\cite{home} of one of  the authors~(M.H.).

\begin{figure}[!ht]
\begin{center}
\vskip-0.4cm
\hspace*{-2cm}\psfig{file=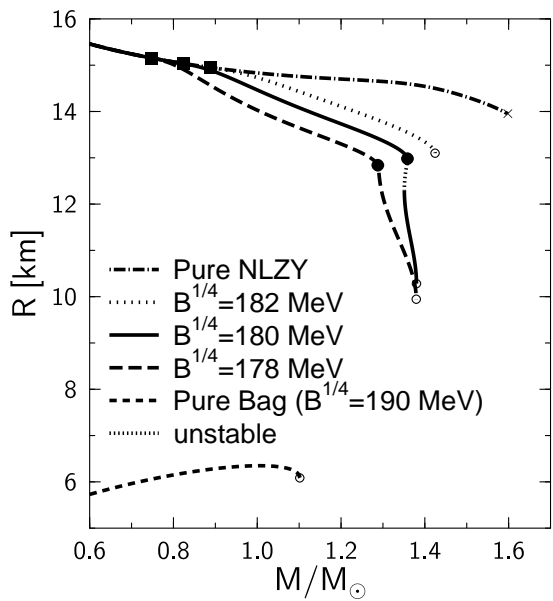,width=0.55\textwidth}
\vskip-4.0cm
\caption{\label{fig:3}
The mass-radius relations for different families of stars. Open 
dots correspond to the maximum mass. The beginning and the end of the 
mixed phase in hybrid stars are shown by full squares and dots.}
\end{center}
\vskip-0.5cm
\end{figure}

In Fig.~\ref{fig:3} we present the mass--radius relations calculated for
several values of $B$ as well as for pure hadronic and quark stars. By open
dots we show the critical configurations with highest possible masses. The pure
hadronic curve are calculated assuming the NLZY EoS at all densities. The part
of this curve corresponding to $M>1.6 M_\odot$ is omitted due to appearance of
densities with negative effective masses of baryons~\cite{scha96}. The
beginning of the MP at the star  center is marked by full squares whereas its
end  and, therefore, the beginning of the pure QP, is marked by full dots.  As
seen in Fig.~\ref{fig:3}, the  model predictions depend strongly on the bag
constant.  Stable hybrid stars are possible only for low bag constants,
$B^{1/4} \leq 180$~MeV. For higher $B$ stars  become unstable even before they
reach high enough density for the formation of a pure deconfined phase in their
interiors. On the other hand, pure quark stars are possible at any $B$\,.

A very interesting feature is found in a narrow interval of $B$ around
$B^{1/4}=180$~MeV, where a second sequence of  stable hybrid stars appears.
Their properties can be  summarized as follows:
\begin{itemize}

\item 
The first sequence of stars can only reach maximum central  baryon densities of
about $0.77$ fm$^{-3}$ which is just at the end of the MP.  As one can see from
Fig.~\ref{fig:3}, stars of this sequence have masses below 
$M_{\rm max}=1.36\,M_\odot$ and radii \mbox{$R> 13$}~km.  
These stars are mainly composed of
hadronic and  MP matter with admixture of the $\Lambda$-hyperons. The
calculation shows that stars near the maximum mass contain a tiny core of pure
quark phase.

\item 
Stars of the second sequence have considerably higher central densities,
$0.9\,{\rm fm}^{-3}<\rho_c<1.53\,{\rm fm}^{-3}$\,. 
Their masses lies in a
narrow interval, $1.35 < M/M_\odot < 1.38$ and their radii are 
noticeably smaller, $10.2\,{\rm km} < R < 12.3\,{\rm km}$. 
These  stars are  mainly composed of quark and MP matter, surrounded 
by a rather thin  layer of HP and the nuclear crust.

\item 
The two sequences of compact stars are separated by unstable region
corresponding to the interval of central densities 
\mbox{$0.78\,{\rm fm}^{-3}<\rho_c<0.9\,{\rm fm}^{-3}$}. 
\end{itemize}

From our more comprehensive study~\cite{han02} involving many other  models of
EoS we conclude that two sequences of compact stars, sometimes called twin
stars, appear only in very exceptional cases. There is no guarantee that this
picture is the one that corresponds to the reality. But we find  it interesting
to study its possible consequences for the dynamics and observable signatures
of compact stars.

\begin{figure}[!ht]
\begin{center}
\centerline{\psfig{file=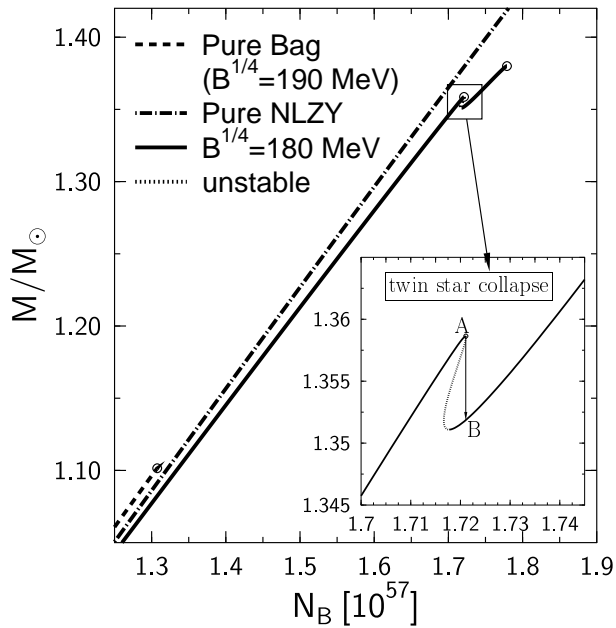,width=0.55\textwidth}}
\vskip-3.2cm
\caption{\label{fig:4}
Mass as a function of baryon number for different families of stars. 
The inset shows the enhanced region of twin stars. The twin star collapse 
corresponds to the transition from A to B.}
\end{center}
\vskip-1.0cm
\end{figure}

\section{The twin star collapse}

One can imagine the situation  when a compact star on the first sequence has a
mass close to $M_{\rm max}$. If this star is a member of binary system, then
its mass and, therefore, its central density may grow due to accretion from a
companion star. Eventually the mass will exceed the maximum value when the
star becomes unstable  with respect to radial compression. Usually it is
assumed that this loss of stability leads to  the collapse into a black
hole. However, our  calculations open another possibility:  the
collapse into the twin star on the second sequence.

Let us assume that no matter is ejected during this process, i.e. the total
baryon number, $N_B$\, is conserved.  Figure ~\ref{fig:4} shows the
gravitational mass of the star as a function of its total baryon number. The
star from the first sequence which reaches the maximum mass (point~A) will
collapse  to its twin star.  The latter is the corresponding star on the second
sequence, i.e. the one which has the same total baryon number (point~B). The
difference in energy between these two stars, $\Delta E$, is given by the
difference in their gravitational masses. In the case considered here, the 
released energy is 
$\Delta E\simeq \mbox{$6\times 10^{-3}\hsp M_\odot$} \simeq 10^{52}$ 
erg. This amount of energy should finally be emitted in one or the other way.

\begin{figure}[!ht]
\begin{center}
\centerline{\psfig{file=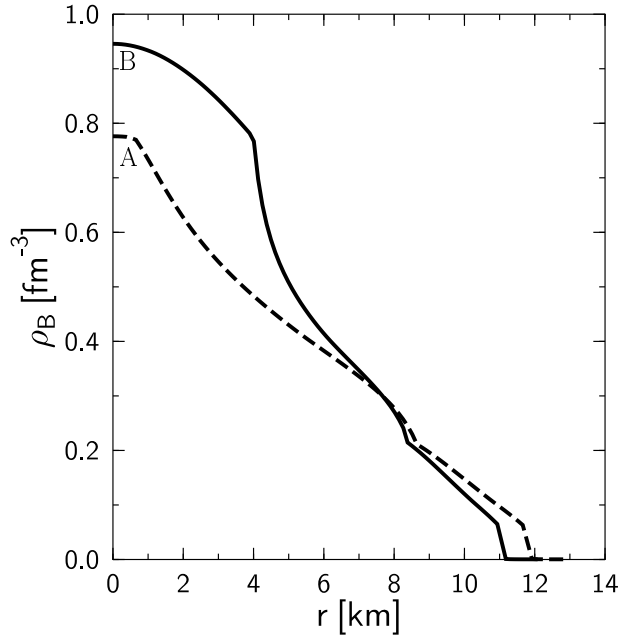,width=0.65\textwidth}}
\vskip-4.1cm
\caption{\label{fig:5}
Baryon density profiles of the twin 
stars A and B as defined in Fig 4.}
\end{center}
\vskip-1.0cm
\end{figure}

The baryonic density  profiles of the twin stars A and B are compared in  
Fig.~\ref{fig:5}. One can see that an extended MP is present in both cases
but  the star B has much larger quark core as compared  to the star A, where
this core is only marginally present.  So, the instability develops practically
at the end of the MP region. In other words, this shows that stars with small
quark   cores are unstable. In the considered example the new equilibrium 
state appears only when the core radius exceeds  about 4 km. This reminds the
well known result from the theory of nonrelativistic stars with a density jump
inside. Namely, if $\rho_1$ and $\rho_2$ are baryon densities just below and
above this jump, then small  dense core becomes unstable if $\rho_1/\rho_2$ is
larger than a certain  critical value (3/2~for incompressible
matter~\cite{ram50}). Of course, in our case there appears no jumps of density,
although its gradients are large near the core boundary.

This kind of instability associated with a first order phase transition  in
hadronic matter was first studied in Ref.~\cite{mig79} and later on in
Refs.~\cite{kam81,hae89}. It was shown that the transition from a small-core 
to a large-core configuration proceeds via large-scale damped  oscillations
around the new equilibrium state. We expect that the   collapse of twin stars
may  proceed in a similar way. After reaching the critical state A the star
looses stability and enters the stage of catastrophic rearrangement. Since the
QP in the core has higher density than the replaced MP, surrounding layers of
the star will acquire collective inward motion. Due to inertial effects, the
star will overshoot the new equilibrated state B and rebound. This will give
rise to the damped oscillations around this state. As simple dimensional
estimates show~\cite{mig79}, the initial implosion and following oscillations
are characterized by the millisecond time scale. 

Such star rearrangement may involve many interesting processes. 
First of all, large
amount of hadronic matter will be transformed into deconfined phase. Since
strangeness and leptonic contents are very different in two phases,  this
transformation will require not only strong-interaction reactions but also weak
processes changing strangeness and lepton numbers.  Such reactions will
inevitably produce large numbers of neutrinos. For instance, to achieve
chemical equilibrium in the deconfined phase one needs reactions  
\mbox{$e+u\to d+\nu$} and \mbox{$u+d\to s+u$} which lead,  respectively, to
emission of neutrinos and production of additional strange quarks.  We believe
that the time scales of these weak processes are much shorter than the
characteristic times of the star collapse.

The total number of neutrinos, $N_\nu$\,, produced during the  transition
process can be estimated as follows. From Fig.~\ref{fig:5} one can see
that  the final star has the quark core with baryon density 
$\rho_{\rm core}\sim 5\hsp\rho_0$ and radius $R_{\rm core}\simeq 4$ km. 
Thus, the core  contains $N_{\rm core}\sim 2\times 10^{56}$ baryons. 
They were initially in the HP at lower density $\rho_B\sim 2\rho_0$\,.
According to Fig.~\ref{fig:2} the fraction of electrons in this 
phase was  about 10\% and practically no electrons were in
the QP. Therefore, the difference between the number of electrons 
in configurations A and~B is $\Delta N_e\sim 2\times 10^{55}$\,
(the direct numerical calculation gives $\Delta N_e = 1.6\times 10^{55}$).
Because of lepton number conservation, all these electrons  should
be transformed into neutrinos and therefore, 
$N_\nu=\Delta N_e\sim 2\times 10^{55}$.
This number is  about 1\% of the total baryon number of the
star $N_B\simeq 1.7\times 10^{57}$\,. The neutrino energies will cover a
broad  range up to a maximum value 
\mbox{$E_\nu^{\,\rm max}=\mu_e+\mu_u-\mu_d$}\,. 
One can estimate $E_\nu^{\,\rm max}$ by assuming that
initially the deconfined phase was produced in the
nonequilibrium state with the flavor composition corresponding to
the~HP. The latter consisted mainly of neutrons, with the quark 
structure $udd$\,, as well as about 10\% admixture of protons and electrons.
Taking $\rho_d\simeq 2\rho_u\simeq 10\rho_0$ and
$\rho_e\simeq 0.5\rho_0$ we get $E_\nu^{\,\rm max}\simeq 150$\, MeV 
which corresponds to the mean neutrino energy $\sim 100$\, MeV.   
Multiplying this energy by $N_\nu$ one obtains that a significant
fraction ($\goo$ 30\%) of the released energy  will be carried away by the
prompt neutrino burst. 

We expect that the remaining energy will be transformed into heat. 
Nonequilibrium processes during the phase transformation
as well as viscosity effects might be responsible for  
dissipation of collective kinetic energy and eventually for
damping of oscillations. Assuming that thermal energy
$\Delta E_T$ is initially dissipated in the quark core, one can
estimate its temperature $T_0$ from the relativistic Fermi gas
formula 
\bel{exef}  
\frac{\pi^2}{4}\frac{T_0^2}{\mu_q}=\frac{\Delta E_T}{3N_{\rm core}}\,,
\ee
where $\mu_q=\mu_b/3\simeq 0.4$ GeV is typical quark chemical potential at
$\rho_B\sim\rho_{\rm core}$\,. For 
\mbox{$\Delta E_T\sim 7 \times 10^{51}$}~erg Eq.~(\ref{exef}) 
gives $T_0\simeq 40$ MeV. At later times the heat wave will propagate through
the whole star and finally will produce photons and, possibly,
electron-positron pairs at its surface. The emission temperature, $T_{\rm
em}$\,, can be estimated  by assuming that the thermal energy is
distributed over the whole stellar matter, which is nonrelativistic outside the
core. Then one obtains $T_{\rm em}\sim 2\Delta E_T/3N_B\sim 2$~MeV.  These
predictions give us a ground to think that the discussed mechanism may serve as
the engine for Gamma-Ray Bursts (GRB)~\cite{pir00}.
Another obvious prediction is that the discussed
rearrangement of the star will lead to a significant change in its 
moment of inertia. In a rotating star this will result in a super-glitch
phenomenon~\cite{maxie}.

\section{Conclusions and outlook}

It is shown that within a realistic hybrid (NLZY-MIT Bag) model two sequences
of compact stars (twin stars) are possible.  We demonstrate that their interiors
differ mainly by the size of pure quark core. The energy difference between two
twin stars with the same baryon number is about $10^{52}$ erg which is
approximately 1\% of their total energy. The transition between twin stars may
be triggered by accretion of mass from a companion star or by some other
processes leading to increase of the central density above the critical value.
After that a catastrophic rearrangement of the star begins. It will first
collapse and then oscillate  around a new equilibrium state.

Our estimates show that in the course of this process  the quark core can be
heated up to temperature of about 40 MeV which is comparable to the one in 
supernova explosions.  Our main prediction is that the  transition between twin
stars  will produce a prompt burst of neutrinos (with energies of about 100
MeV) followed by a gamma ray burst (with photon energies of about 1 MeV).
These features of the twin star collapse make it a
potential candidate for the GRBs. On the other hand, there are
some characteristics of GRBs, e.g. beaming of radiation, which is impossible to
explain without additional assumptions invoking rotation and strong magnetic
fields. This problem needs further study.

It is interesting to note that the presented model  predicts maximum masses of
compact stars, $(1.35-1.38) M_\odot$\,,  which agree well with observed
masses of pulsars~\cite{gle97}.  Moreover, the
twin stars of the second sequence have nearly the same masses in the broad
range of~$\rho_c$\,. If pulsars have quark cores,  
this may explain why their measured masses are
clustered near the mass of about $1.4 M_\odot$\,. 

Other phase transitions may take place in the interiors of compact stars, e.g.
kaon condensation, color superconductivity etc. In Ref.~\cite {scha02} it was
shown that formation of  metastable hyperonic matter can also lead  (in a
certain range of parameters) to the twin star solutions.  The comparison of
twin star properties predicted by different models  will be given in a separate
work~\cite{han02}.

\section*{Acknowledgements}
The authors thank J. Schaffner--Bielich for fruitful discussions.
This work was funded in part by  GSI, DAAD, DFG	 
and the Hessische  Landesgraduiertenf\"orderung.  
A.B. thanks the Alexander von Humboldt Foundation for support.
I.N.M. and L.M.S. acknowledge support from the RFBR Grant No.~00--15--96590.

\end{document}